\documentclass[epj,referee]{svjour}
\usepackage{graphics}
\usepackage{amssymb}
\begin{document}
\hugehead

\title{Nuclear Polarization of Molecular Hydrogen Recombined on 
a Non-metallic Surface}

\author{ 
The HERMES Collaboration \medskip \\
A. Airapetian,$^{30}$
N. Akopov,$^{30}$
Z. Akopov,$^{30}$
M. Amarian,$^{6,30}$
V.V. Ammosov,$^{22}$
A. Andrus,$^{15}$
E.C. Aschenauer,$^{6}$
W. Augustyniak,$^{29}$
R. Avakian,$^{30}$
A. Avetissian,$^{30}$
% ...
}

\institute{ 
$^1$Department of Physics, University of Alberta, Edmonton, Alberta T6G 2J1, Canada\\
$^2$Physics Division, Argonne National Laboratory, Argonne, Illinois 60439-4843, USA\\
$^3$Istituto Nazionale di Fisica Nucleare, Sezione di Bari, 70124 Bari, Italy\\
$^4$Nuclear Physics Laboratory, University of Colorado, Boulder, Colorado 80309-0446, USA\\
$^5$DESY, Deutsches Elektronen-Synchrotron, 22603 Hamburg, Germany\\
$^6$DESY Zeuthen, 15738 Zeuthen, Germany\\
$^7$Joint Institute for Nuclear Research, 141980 Dubna, Russia\\
$^8$Physikalisches Institut, Universit\"at Erlangen-N\"urnberg, 91058 Erlangen, Germany\\
$^9$Istituto Nazionale di Fisica Nucleare, Sezione di Ferrara and Dipartimento di Fisica, Universit\`a di Ferrara, 44100 Ferrara, Italy\\
$^{10}$Istituto Nazionale di Fisica Nucleare, Laboratori Nazionali di Frascati, 00044 Frascati, Italy\\
$^{11}$Fakult\"at f\"ur Physik, Universit\"at Freiburg, 79104 Freiburg, Germany\\
$^{12}$Department of Subatomic and Radiation Physics, University of Gent, 9000 Gent, Belgium\\
$^{13}$Physikalisches Institut, Universit\"at Gie{\ss}en, 35392 Gie{\ss}en, Germany\\
$^{14}$Department of Physics and Astronomy, University of Glasgow, Glasgow G12 8QQ, United Kingdom\\
$^{15}$Department of Physics, University of Illinois, Urbana, Illinois 61801-3080, USA\\
$^{16}$Laboratory for Nuclear Science, Massachusetts Institute of Technology, Cambridge, Massachusetts 02139, USA\\
$^{17}$Randall Laboratory of Physics, University of Michigan, Ann Arbor, Michigan 48109-1120, USA \\
$^{18}$Lebedev Physical Institute, 117924 Moscow, Russia\\
$^{19}$Sektion Physik, Universit\"at M\"unchen, 85748 Garching, Germany\\
$^{20}$Nationaal Instituut voor Kernfysica en Hoge-Energiefysica (NIKHEF), 1009 DB Amsterdam, The Netherlands\\
$^{21}$Petersburg Nuclear Physics Institute, St. Petersburg, Gatchina, 188350 Russia\\
$^{22}$Institute for High Energy Physics, Protvino, Moscow region, 142281 Russia\\
$^{23}$Institut f\"ur Theoretische Physik, Universit\"at Regensburg, 93040 Regensburg, Germany\\
$^{24}$Istituto Nazionale di Fisica Nucleare, Sezione Roma 1, Gruppo Sanit\`a and Physics Laboratory, Istituto Superiore di Sanit\`a, 00161 Roma, Italy\\
$^{25}$Department of Physics, Simon Fraser University, Burnaby, British Columbia V5A 1S6, Canada\\
$^{26}$TRIUMF, Vancouver, British Columbia V6T 2A3, Canada\\
$^{27}$Department of Physics, Tokyo Institute of Technology, Tokyo 152, Japan\\
$^{28}$Department of Physics and Astronomy, Vrije Universiteit, 1081 HV Amsterdam, The Netherlands\\
$^{29}$Andrzej Soltan Institute for Nuclear Studies, 00-689 Warsaw, Poland\\
$^{30}$Yerevan Physics Institute, 375036 Yerevan, Armenia\\
$^{31}$Department of Physics, University of Wisconsin-Madison, Madison, Wisconsin 53706 USA\\
$^{32}$Physics Department, University of Liverpool, Liverpool L69 7ZE, United Kingdom\\
}
\date{Received: \today / Revised version:}

\titlerunning{Nuclear Polarization of Molecular Hydrogen}
\authorrunning{The HERMES Collaboration}

\abstract{The nuclear polarization of $\mathrm{H}_2$ molecules formed by 
recombination of 
nuclear polarized H atoms on the surface of a storage cell 
initially coated with a silicon-based
polymer has been measured by using
the longitudinal double-spin asymmetry in deep-inelastic positron-proton 
scattering. The molecules are found to have a substantial nuclear polarization,
which is evidence that initially polarized atoms retain their nuclear
polarization when absorbed on this type of surface.
} 
\maketitle

\section{Introduction}

During recent years, increased use has been made of nuclear-polarized 
hydrogen and
deuterium gas targets, which are placed in the circulating beams of storage
rings. 
In order to increase their thickness over that obtained 
by a jet of polarized atoms, the beam from atomic sources is directed
into an open-ended, cooled tube ({\it{storage cell}}) 
in which the atoms make several hundred wall collisions before escaping.
In order to maintain high polarization of the atomic gas, the storage cells 
are usually coated 
with {\em e.g.} chemically nonreactive polymers \cite{Dri0,Dri1,Dri2}.
Examples of the successful
use of this technique are measurements of spin correlation 
parameters in 
proton-proton scattering at the Indiana University Cyclotron Facility (IUCF)
\cite{IUCF}, studies of nucleon electromagnetic form factors at the storage
rings VEPP-3 in Novosibirsk 
\cite{Novosibirsk} and AmPS at NIKHEF in Amsterdam \cite{NIKHEF}, and
deep inelastic scattering of positrons off polarized H nuclei at the HERMES 
experiment situated in the HERA storage ring at DESY in Hamburg 
\cite{HERMES,spectrom}.

A precise knowledge of the nuclear polarization of the target is required
to extract interesting physics quantities from measured polarization 
asymmetries \cite{IUCF,Novosibirsk,NIKHEF,spectrom}. 
At the present time, there are no convenient known scattering
processes for the high-energy positrons beams available at HERA, 
that could provide a measure of the absolute value
of the target polarization with 
acceptable statistics. 
For the HERMES experiment, therefore, a target polarimeter 
with good absolute precision
has been constructed \cite{BRP}.
However, while the polarization of a sample of atoms from the cell is
measured directly, the nuclear polarization of $\mathrm{H}_2$ molecules which
originate from recombination of hydrogen atoms in the target cell is unknown. 

The nuclear polarization of recombined $\mathrm{H}_{2}$ molecules was
recently measured in a separate experiment at IUCF \cite{TOM}. 
The nuclear polarization of the molecules obtained by recombination of
polarized atoms on a copper surface was measured by elastic 
proton-proton scattering.
The other existing measurement of polarization of recombined atoms 
concerns tensor polarized 
D atoms recombining on a copper surface at AmPS \cite{Jo}.
The absence of any measurement
of the nuclear polarization of molecules recombined on a surface 
involving a silicon-based polymer motivated the measurement described
in the present paper.
The measurement was performed by comparing the double-spin 
asymmetries observed in deep inelastic scattering of polarized positrons from
gaseous polarized hydrogen target, while the fraction of molecules was
varied.
The experiment was carried out using the internal target and spectrometer of 
the HERMES experiment at HERA. The observation of a significant nuclear 
polarization
in the recombined molecules could be interpreted as evidence for the existence
of long-lived nuclear polarization of atoms chemisorbed onto an insulating
surface.

\section{The HERMES Hydrogen Target}

\subsection{Setup}

\begin{figure}
\resizebox{0.47\textwidth}{!}{
\includegraphics{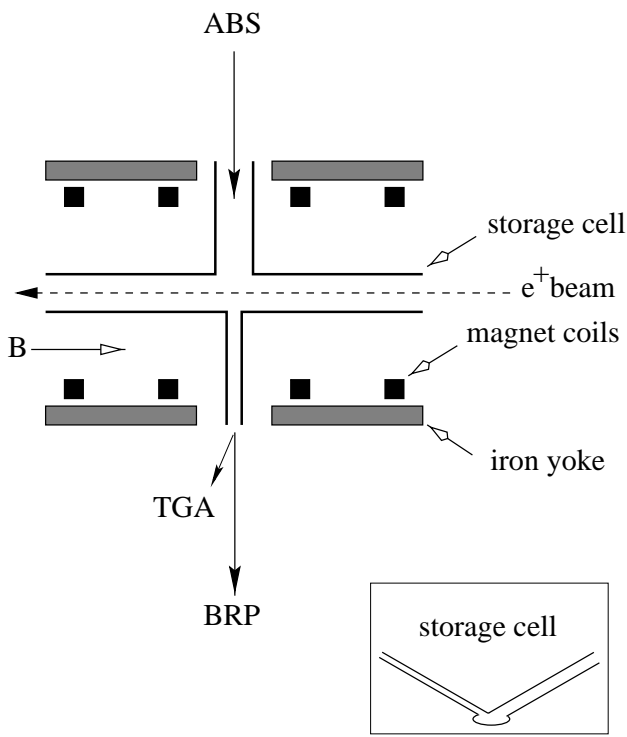}
}
\caption[]{A schematic layout of the HERMES hydrogen target.
From left to right: injection tube for 
atomic beam source (ABS), target chamber with cell
and magnet, sample tube taking gas to
target gas analyzer (TGA) and Breit-Rabi polarimeter (BRP).}
\label{fig_target}
\end{figure}

A schematic diagram of the HERMES target is shown in Fig.~\ref{fig_target}.
A beam of nuclear-polarized hydrogen atoms is generated by an 
atomic beam source 
(ABS) \cite {nass} and injected into the center of the storage
cell \cite{pfd} via a side tube. The atoms then diffuse to
 the open ends of the
cell where they are removed by a high speed pumping system. 
A magnetic field of 330 mT surrounding the cell
provides a quantization axis for the spins and inhibits
nuclear spin relaxation by effectively decoupling nucleon and electron spins.
The atoms diffusing
from a second side tube of the cell
are analyzed by a Breit-Rabi polarimeter (BRP)
\cite{BRP} which
measures the polarization of the atoms and by a target gas 
analyzer (TGA) \cite{TGA} which determines
the atomic fraction. 

\subsubsection{Cell surface}

The storage cell is constructed from 99.5\% pure aluminum, 75 $\mu$m thick, it
has an elliptical cross section with 29 mm $\times$ 9.8 mm axes and a length of
400 mm.
The maximum density in the cell is about 
$\mathrm{4 \times 10^{12}}$ $\mathrm{atoms/cm^{3}}$.
The cell surface
was initially coated with {\it{Drifilm}}, a silicon-based polymer
which is radiation hard,
as required in an accelerator environment \cite{Dri3}.
The coating procedure
is described in detail in Ref. \cite{Dri2}; the bake-out step of the 
procedure was done at 220 $^\circ \mathrm{C}$ for 12 hours. 
The uniformity of the coating was tested visually.
After coating, the cell was
stored in air before being installed in the scattering chamber in HERA.
The pressure in this chamber is in the $10^{-9}/10^{-8}$ mbar range 
and the 
cell was kept at a temperature of 100 K. The storage cell is protected 
from the intense
synchrotron radiation of the positron beam
by a system of tungsten collimators upstream of the target. 
However, the interior surface of the target cell is intensely irradiated
by positive ions that are produced in the gas by the beam and then 
stochastically accelerated (``heated'') by the electric field of the very
compact beam bunches as the ions spiral in the magnetic holding field until
they attain an orbit large enough to reach the cell wall with energies of
the order of 100 eV. Monte Carlo simulations of this process result in 
estimates of the dose to the cell wall surface around 10 Gy/s.
After exposing the coated target cell in the proximity of the positron beam
of the HERA storage ring the quality of the surface is tested again with
the water-bead method \cite{Dri0,Dri1,Dri2}, which was also used after
production of the cell. It is routinely observed that the original 
hydrophobic Drifilm surface become hydrophilic as indicated by the changes
in the wetting properties shown in this test.
The measurement described in this work was carried out
when the cell
had been exposed to radiation for 4 months, and spanned a period of 3 weeks.

\subsection{Target polarization}

The molecules inside the HERMES target have three origins: 
residual gas in the scattering chamber containing the storage cell, 
molecules ballistically injected by the ABS
and atoms that recombine after entering the cell.
The hydrogen molecules present in the residual gas of the chamber
originate from atoms recombining the surface of the target chamber itself,
where they rapidly lose their polarization.
The ballistically injected molecules were never dissociated in the
beam source and therefore never polarized. 
For these reasons the nucleons belonging to residual gas and ballistically
injected molecules are assumed to be not polarized,
while the recombined molecules
may have a remnant nuclear polarization. 

Each contribution is directly measured with specific calibrations
using the TGA as described in Ref. \cite{TGA}.


The atomic fraction $\alpha$ is defined as:
\begin{equation} 
\alpha=\frac{n_{a}}{n_{a}+2n_{m}},
\end{equation}
where $n_a$ and $n_m$ are the atomic and molecular populations averaged 
over the cell.
In order to distinguish unpolarized and potentially polarized molecules, the 
atomic fraction $\alpha$ is separated into two factors:
$\alpha_0$ and $\alpha_r$; $\alpha_0$ defines the atomic fraction 
in absence of recombination,
while $\alpha_r$ defines the fraction of atoms surviving recombination
in the storage cell.

The measured target polarization $P_T$
as seen by the positron beam is described by the following expression: 
\begin{equation} 
P_T =\alpha_{0}\alpha_{r}P_{a} + \alpha_{0}(1-\alpha_{r})\beta P_{a}.\\
\label{P}
\end{equation}
Here $P_{a}$ is the nuclear polarization 
of the atoms in the cell. The parameter
$\beta=P_{m}/P_{a}$
represents the relative nuclear polarization of the
recombined molecules 
with respect to the atomic polarization before 
recombination ($0 \le \beta \le 1$), and it is the subject of
the measurement presented in this paper.

As the particles sampled from the center of the cell have
experienced a different average number of wall collisions than the ones 
at the ends of the cell, the properties of the measured sample are 
slightly different from the average values in the cell. 
In order to relate the average values over the cell
to the corresponding measurements on the sample beam,
{\it{sampling corrections}} have to be introduced.
The sampling corrections  
depend on the sensitivity of the BRP
and the TGA to the various parts of the cell and are derived from
Monte-Carlo simulations of the molecular flow of gas particles
traveling through the storage cell as described in Ref. \cite{SAM}.
The corrections and their uncertainties take extreme assumptions of
cell surface quality into account and constitute the dominant
contribution to the systematic error in the target polarization.

\section{Measurement}

\subsection{The HERMES Spectrometer}

The HERMES experiment is installed in the HERA storage ring at 
the DESY laboratory in Hamburg. It uses the 27.6 GeV positron beam
 with a typical injected current of 40 mA. 
The positrons are transversely polarized
by emission of spin-flip synchrotron radiation \cite{Sokolov}. 
Longitudinal polarization
of the positron beam at the interaction point is achieved by spin rotators
situated upstream and downstream of the experiment. The beam polarization
is continuously measured using Compton back-scattering of circularly polarized
light. Two polarimeters with low
systematic uncertainties, are used, one measuring the transverse polarization
in the HERA West straight section \cite{Barber} and the other measuring the
longitudinal polarization near the HERMES target \cite{lpol}.
Positron identification in the momentum range $2.1$ to $27.6$ GeV is 
accomplished by using a lead-glass calorimeter wall, a transition-radiation detector and a preshower 
hodoscope. This system provides positron identification with an average 
efficiency higher than 
$98$ \% and a negligible hadron contamination. The luminosity
is measured by detecting Bhabha scattered target electrons in coincidence
with the scattered positrons in a pair of electromagnetic calorimeters
\cite{lumi}.
The HERMES spectrometer is described in detail in Ref. \cite{spectrom}.

\subsection{Method}
The method adopted to extract the relative molecular polarization $\beta$
makes use of the
double spin asymmetries observed in deep-inelastic scattering of
longitudinally polarized positrons off
a longitudinally polarized proton target.
The cross sections $\sigma^+$ and 
$\sigma^-$ for parallel and antiparallel relative orientations of
beam and target polarizations,
are related to the unpolarized cross section $\sigma_0$ by:
\begin{equation}
\sigma^{+/-} = \sigma_0 (1\mp P_B P_T A_{||}),
\end{equation}
where $P_B$ and $P_T$ are the beam and target polarizations and 
$A_{||}$ is the experimental double spin cross section asymmetry.
The actually measured count rate asymmetry $C_{||}$, however, is smaller.
It is given by:
\begin{equation}
C_{||} = \frac{(N/L)^- - (N/L)^+}{(N/L)^- + (N/L)^+} = P_B P_T A_{||},
\label{C}
\end{equation}
where $(N/L)^{+,-}$ denotes the number of events with parallel 
(anti-parallel) beam and target spins orientations
corrected for the background, normalized to the corresponding 
luminosity. 

Two measurements have been performed 
to determine the polarization of the molecules;
one with an almost purely atomic target
of known polarization
and a second with an enhanced molecular fraction (or smaller value of 
$\alpha_r$). 
The increase in the amount of recombined molecules
was achieved by increasing the temperature of the target cell 
from 100 K (normal running conditions) to
260 K. Over this temperature range the main mechanism
thought to be responsible for recombination in the target cell
is the Eley-Rideal mechanism \cite{ER} 
in which an atom
in the gas phase hits a chemically bound atom on the surface with enough
kinetic energy to overcome the activation barrier.
The increase in the target temperature results in an increase of the
kinetic energy of the atoms in the volume and in a higher recombination
probability. No recombination happens on physically adsorbed atoms,
as their coverage is negligible at 260 K. Detailed studies on 
various aspects of the recombination process occurring
at the HERMES cell surface are reported in Ref. \cite{rec}.

The target parameters for the two different 
temperatures are given in Table \ref{tab:a}: a large increase
in the fraction ($1-\alpha_r$) of recombined atoms 
from 0.055 at 100 K to 0.74 at 260 K is observed.
As the double spin asymmetry  $A_{||}$ does not depend
on target or beam polarization, we can write:
\begin{equation}
(\frac{C_{||}}{P_T P_B})_{100 K} =
(\frac{C_{||}}{P_T P_B})_{260 K}.
\label{Apar}
\end{equation}
The target polarizations are given by (see Eq. \ref{P}):
\begin{eqnarray}
P_T^{100 K}=\alpha_0^{100 K}[\alpha_r^{100 K}+
(1-\alpha_r^{100 K}){\beta^{100 K}}]P_a^{100 K}\\
P_T^{260 K}=\alpha_0^{260 K}[\alpha_r^{260 K}+
(1-\alpha_r^{260 K}){\beta^{260 K}}]P_a^{260 K}.
\end{eqnarray}
The values for $\alpha_0^{100 K, 260 K}, \alpha_r^{100 K, 260 K},
P_a^{100 K, 260 K}$ are reported in  Table \ref{tab:a}.

\begin{table}
\caption{Atomic polarization and atomic fraction as measured
at 100 K and 260 K in the HERMES storage cell, for the selected data sample.}
\label{tab:a}       
\begin{tabular}{llll}
\hline\noalign{\smallskip}
$T_{cell}$ & $P_a$ & $\alpha_0$ & $\alpha_r$  \\
\noalign{\smallskip}\hline\noalign{\smallskip}
100 K & $0.906 \pm 0.01$ & $0.96 \pm0.03$ & $0.945 \pm 0.035$\\
260 K & $0.939 \pm 0.015$ & $0.96 \pm0.03$ & $0.26 \pm 0.04$\\
\noalign{\smallskip}\hline
\end{tabular}
\end{table}

\subsection{Extraction of the parameter $\beta$}
In order to extract information on $\beta$, a 
minimization procedure has been used. 
Equation \ref{Apar} has two unknowns: $\beta^{100 K}$ and $\beta^{260 K}$,
which are independent since the surface conditions
at 100 K and 260 K are assumed to be different.
The value
$\beta^{260 K}$ has been determined by varying its value until the function 
$\chi^{2}(\beta^{260 K})$ described by: 
\begin{equation}
\chi^{2}(\beta^{260 K}) = \sum_{i} \left[ \frac
{(\frac{C_{||i}}
{P_{B}P_{T}})_{260 K}
-(\frac{C_{||i}}
{P_{B}P_{T}})_{100 K}}
{\left( (\frac{\delta_{C||i}}{P_{B}P_{T}})^2_{260 K} 
+ (\frac{\delta_{C||i}}{P_{B}P_{T}})^2_{100 K} \right)^{1/2}
} \right]^2
\label{F}
\end{equation}
was minimized. The minimization has been studied as a function of the parameter
$\beta^{100 K}$. 
In Eq. \ref{F} $C_{||i}$ is the count rate asymmetry in the
$i^{th}$-bin of the kinematic plane and
$\delta_{C||i}$ is its associated statistical uncertainty. 
(The statistical uncertainties of $P_T$ and $P_B$ are negligible here).
The summation is made over all the kinematic bins of the
asymmetry measurement. 
The binning of the kinematic plane is the same as
the one adopted in the extraction of the $g_1^{p}$ spin-dependent
structure function of the proton from
the $A_{||}$ asymmetry, as described in \cite{g1}.
It was verified, by solving Eq. \ref{Apar} for all bins separately, that
$\beta$ does not depend on the bin number and therefore Eq. \ref{F} gives
the best possible result. 

As at 100 K only a few percent of the gas is in molecular 
form ($1-\alpha_r = 0.055$, as can be seen from
Table \ref{tab:a}), the final result is insensitive to the value
assumed for $\beta^{100 K}$. Using the conservative assumption
$0 \le \beta^{100 K} \le 1$, 
the following value has been obtained for $\beta^{260 K}$:
\begin{equation}
\beta^{260 K} = 0.68 \pm 0.09_{stat} \pm 0.06_{syst}.
\label{beta}
\end{equation}

The uncertainty of the measurement is dominated by the low statistics
of the data taken at 260 K (0.19 million events) compared
to the 100 K data sample (2.5 million events).
The statistical uncertainty 
corresponds to the change in the value of $\chi^{2}(\beta^{260 K})$ by one unit
from its minimum, and was found to be independent of the value of 
$\beta^{100 K}$.
The main source of the systematic uncertainty arises from the knowledge of
$\alpha_r^{260 K}$. Unfortunately, the sampling corrections described
in Section 2.2 cannot be applied at 260 K as the associated systematic
uncertainty grows rapidly if $\alpha_r$ is significantly below unity.
Hence, an alternative method has been used to correct for this effect.
Coincident Bhabha-scattering event rates
measured by the 
luminosity monitor are
proportional to the target density, which depends on 
the atomic fraction. 
(The molecules have an average velocity lower by a factor $\sqrt{2}$).
The comparison of the
Bhabha rates detected at 100 K and 260 K with constant atomic flux
injected in the cell allows one to limit the interval of 
$\alpha_r^{260 K}$  to the range reported in Table \ref{tab:a}.
Some systematic uncertainties, such as those in $\alpha_0$ and $P_B$,
are common to both the 100 K and the 260 K measurements 
and cancel in the ratio in Eq. \ref{F}.

\begin{figure}
\resizebox{0.5\textwidth}{!}{
\includegraphics{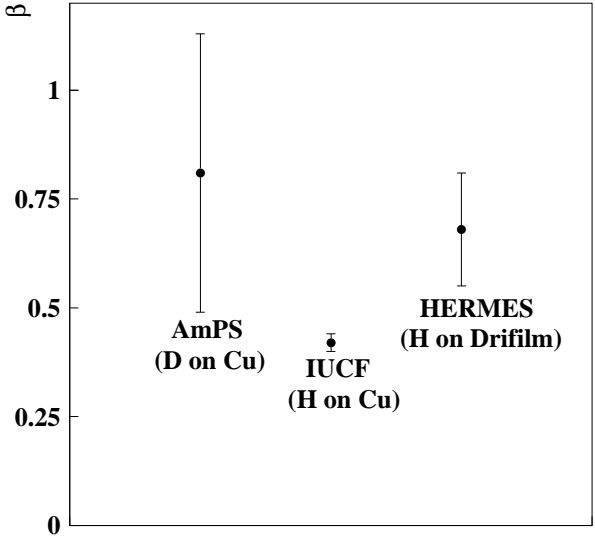}
}
\caption[]{Summary of the existing measurements of the nuclear polarization
of recombined molecules.
The newly obtained HERMES measurement at 260 K 
with a holding field of 330 mT is compared to the measurement by AmPS and 
IUCF obtained at room temperature and 
magnetic holding fields of 28 mT and 440 mT respectively.}
\label{sumbeta}
\end{figure}

In Figure \ref{sumbeta} the result of the present studies on the nuclear 
polarization of recombined H molecules are compared to previous measurements.
In each case a non-zero nuclear polarization is observed, 
but the different surface conditions used in each experiment
preclude a more quantitative comparison of the values themselves. 
Nevertheless, in the discussion which follows, an explanation of these 
differences in terms of the variation in the properties of the surfaces 
has been attempted.

\section{Discussion}

\subsection{Nuclear polarization of surface atoms}

By assuming that the nucleon spins are not affected by the
recombination process,
the nuclear polarization of the molecule at its formation ($P^0_m$) 
can be evaluated by taking the average value of the 
polarization of an atom coming from
the volume ($P{_a}$) and one resident on the surface ($P{_s}$):

\begin{equation}
P^0_{m}={\frac{P_{a}+P_{s}}{2}}.
\label{P:m}
\end{equation}

The loss of polarization of the molecule after recombination has been
described in Ref. \cite{TOM}. 
In free flight, the internal molecular fields $B_c$ from the spin-rotation 
interaction and the direct dipole-dipole interaction
cause the nuclei to rapidly precess around a direction which is skew to the
external field by $B_c/B$. The orientation of $B_c$ is randomized at each
wall collision. Between successive wall collisions, the component of the 
polarization along the external field decreases by an amount $(B_c/B)^2$
and after $n$ wall bounces:

\begin{equation}
P_m = P^0_{m} e^{-n(B_{c}/B)^2},
\label{P:0}
\end{equation}
where $B_c$ for $\mathrm{H}_2$ is 6.1 mT \cite{TOM}. 
For the HERMES cell, we have the value $B \approx 330$ mT. 
Moreover, it has been estimated \cite{SAM} that the number of wall collisions
is of the order of 300. Hence, using Eq. \ref{P:0}, it follows that 
$P_m \approx 0.9 P^0_m$.
By taking the extracted value of $\beta^{260 K}$ and the measured value 
of $P_a^{260 K}$ from Table \ref{tab:a}, we are able to give an
estimate for the polarization of the atoms on the surface making use of 
Eq. \ref{P:m}:
\begin{equation}
P_s^{260 K} = 0.46 \pm 0.22_{tot}.
\label{P_s}
\end{equation}

The total uncertainty results from adding the statistical and systematic 
uncertainties in quadrature.
This non-zero result provides evidence that chemically bound surface 
atoms appear to retain some nuclear polarization. 

\subsection{Relaxation and Dwell Times}

The present result
differs from that of the experiment described in 
Ref. \cite{TOM}, where the measured value of $\beta$ lower than 0.5 
is compatible with a zero residual polarization of the atoms on a copper
surface. 
As both measurements are sensitive to the nuclear polarization of the 
molecular gas coming from the recombination process and one atom of each
recombined molecule comes from the surface, they actually compare 
the nuclear spin relaxation time $\tau_r$
of atoms on the surface with the dwell time $\tau_d$ that the atoms
reside on the surface. 
These times are related by the equation \footnote{The expression has been 
determined by assuming the atomic polarization of the surface atoms
to decay exponentially with a 
characteristic time $\tau_r$ and integrating over an exponential distribution
for the dwell times with characteristic time $\tau_d$.}:
\begin{equation}
P_s = P_a \frac{\tau_r}{\tau_d + \tau_r}.
\label{tau_d}
\end{equation}
By using the result of Ref. \cite{TOM} and those presented in the previous
section, it is concluded that $\tau_r \approx \tau_d$ for the HERMES cell
surface and
$\tau_r \ll \tau_d$ for the copper surface, 
suggesting that, as expected, on a bare metal surface the relaxation processes
are much stronger than on an inert surface like silane.

An upper limit for the dwell time
$\tau_d$ can be derived by considering a uniform surface where all
sites are recombining and by using the inequality:
\begin{equation}
\tau_d < \frac{N_s}{N_{inj}(1 - \alpha_{r})},
\label{tau_r}
\end{equation}
where $N_s$ is the total number of catalytic sites on the cell
surface S and $N_{inj}$ is the number of injected atoms per second.
By assuming an area per surface site of 4$\mathrm{\AA}^2$, taking
$S \approx 0.026$ $\mathrm{m}^{2}$ \cite{pfd},
$N_{inj} \approx 6.5 \times 10^{16}$ atoms/s \cite{nass} and 
$\alpha_r \approx 0.26$ (Tab. \ref{tab:a})
one obtains $\tau_d \lesssim 14$ s and a similar limit for the relaxation time
$\tau_r$.

\section{Conclusions}
In summary, the longitudinal double spin asymmetry in deep-inelastic 
positron-proton scattering
has been used to measure for the first time the nuclear 
polarization of the molecules produced by 
recombination of hydrogen atoms on a storage cell initially prepared
with Drifilm coating.
The measurement shows that the molecules emerging from recombination
on this surface retain a large degree of polarization. In absence
of a polarizing mechanism after leaving the wall, significant polarization
can be inferred for the atoms on the wall. The application of a
simple model allowed to derive an estimate for the depolarization time of
the atoms on the surface.

The present finding of a non-negligible nuclear polarization retained
by the atoms on a surface prepared with Drifilm is encouraging in view 
of hopes for the 
developing of a nuclear polarized gaseous molecular target.

\section{Acknowledgments}
We gratefully acknowledge the DESY management for its support and the DESY 
staff and the staff of the collaborating institutions. This work was 
supported by the FWO-Flanders, Belgium; the INTAS contribution from the 
European Commission; the European Commission IHP program under contract
HPRN-CT-2000-00130; the German Bundesministerium f\"ur Bildung und
Forschung (BMBF); the Italian Istituto Nazionale di Fisica Nucleare (INFN);
Monbusho International Scientific Research Program, JSPS, and Toray Science
Foundation of Japan; the Dutch Foundation for Fundamenteel Onderzoek der
Materie (FOM); the U.K. Particle Physics and Astronomy Research Council and
Engineering and Physical Sciences Research Council; and the U.S. 
Department of Energy and National Science Foundation.

\end{document}